# Printed Circuit Board Based Rotating Coils for Measuring Sextupole Magnets

J. DiMarco

*Abstract*—The use of Printed Circuit Boards (PCBs) for the inductive pick-up windings of rotating coil probes has made the construction of these precision magnetic measurement devices much more accessible. This paper discusses the design details for PCBs which on each layer of the board provide for simultaneous analog bucking (suppression) of dipole, quadrupole, and sextupole field components so as to more accurately measure the higher order harmonic fields in sextupole magnets. Techniques to generate designs are discussed, as well as trade-offs to optimize sensitivity. Examples of recent sextupole PCBs and their performance are given.

## I. INTRODUCTION

The rotating coils designed for magnetic measurements rely on precise windings to obtain accurate measurements of harmonic fields. An important component of this is satisfactory suppression (i.e., 'bucking') of main field components, so that the coils mitigate against spurious harmonics, which can be generated if coil rotation is imperfect, such as having transverse or torsional vibrations [1]. Because of the high precision involved in the manufacture of Printed Circuit Boards (PCBs) (trace placement on a layer at the level of microns), rotating coil probes employing PCBs can achieve high levels of bucking. The so-called 'bucking ratio' (the ratio of main field signal before/after bucking) can typically be several hundred for dipole and quadrupole probes built in this way [2]. The first probes suppressing dipole, quadrupole and sextupole fields (DQS bucking) [3] were built with a design that was not ideal, since it involved other layers of the PCB, and the relative trace placement accuracy of a PCB is most advantageous on a single layer. To use a single layer also means multiple layers go into boosting signal size, not merely bucking. This paper outlines the design of PCBs for magnetic measurement probes which feature DQS bucking on each single layer. The design of the probes is reviewed in detail, and features which can help optimize the DQS-buck PCB circuits are discussed. Examples of fabricated probes are also given.

## II. QUADRUPOLE BUCKING CIRCUIT

The standard way to calculate sensitivity of a PCB circuit is to account for the contribution of each wire according to

$$K_n = \sum_{j=1}^{N_{wires}} \frac{L_j R}{n} \left( \frac{(x_j + i y_j)}{R} \right)^n (-1)^j \quad (1)$$

Here $L$ is the length of a given wire and $R$ is the reference radius. The $(-1)^j$ gives the sign of the current flow of each wire and the $(x_j, y_j)$ are the locations of the wires with respect to the rotation axis. Typically, PCB probe layouts are designed with fixed grid spacing – this is easier for designers, and less susceptible to round-off errors which can inadvertently shift traces – and these will be used in the examples here.

A sample vertical representation of a standard PCB layer cross-section for radial bucking of dipole and quadrupole fields is shown in Fig. 1. In this view, the '+1' represents the centroid of wires traveling axially 'positive' over the length of the PCB, and those with '-1', returning. The radial extent of the PCB is shown by wires at X=0, and a Y value given in the first column, with probe rotation being around [0,0]. The locations of wires are in mm for the examples shown.

Fig. 1. A representative view of a PCB cross-section, positioned vertically, with wire loop tracks, running in/out the page. Rotation center is at Y=0.

There are 4 sets of loops or 'tracks' seen in the cross-section. Pairs of identical windings separated in radius with opposite chirality, such as T1 + T3 or T2 + T4 will buck the dipole field (because the dipole does not have radial dependence), each forming gradient coils. When these two pairs, having opposite chirality to each other, are similarly combined, the gradient that each of them measures is also bucked, leaving a DQBuck signal [2].

Practically, some space at the center of each track is necessary since the loop spirals to the center as it winds inward and must have a 'via' in the PCB to go to the opposite layer, where it can 'unwind' in symmetric fashion. Since each layer has its own bucking, stacking misalignment of the various layers has little impact on probe quality [2].

## III. COMPACT WINDING QUADRUPOLE BUCKING

To make the DQ-buck circuit more compact, the distance between T1/T3 and T2/T4 can be changed – as long as they are the same, the DQ-bucking is preserved. Fig. 2 shows that moving both T3 and T4 to 'overlap' in position with the other tracks allows a more compact DQBuck design, with the resultant shown in Fig. 3.

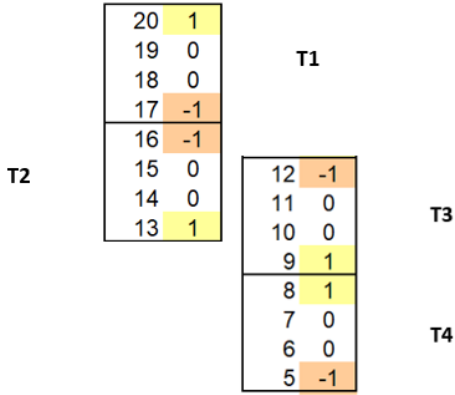

Fig. 2. Repositioning T3 and T4 to make a more compact DQBuck design.

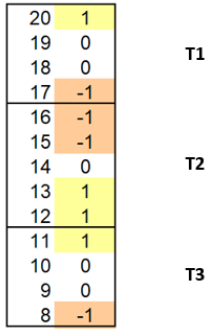

Fig. 3. Compact DQBuck design configuration example.

Note that in Fig. 3, the T2 center is half-way between the centers of T1 and T3, and that the effective weighting at the same track width as T1/T3 is a factor 2 larger and of opposite sign (i.e. T1 and T3 have a width of 3, and the T2 centroid width is also 3, 12.5 to 15.5, but with 2 turns instead of 1). This suggests that we can treat the winding as a group, taking the centroid of the conductors and weighting it by the number of conductors in the group to determine the bucking condition. Addressing this numerically, from Eqn. (1) and Fig. 4,

$$K_2 = N_1\left[\left(x+\frac{w_1}{2}\right)^2 - \left(x-\frac{w_1}{2}\right)^2\right]$$
$$- N_2\left[\left(x+d+\frac{w_2}{2}\right)^2 - \left(x+d-\frac{w_2}{2}\right)^2\right]$$
$$+ N_1\left[\left(x+2d+\frac{w_1}{2}\right)^2 - \left(x+2d-\frac{w_1}{2}\right)^2\right]$$
$$= 4x*(2w_1N_1 - w_2N_2) + 4d*(2w_1N_1 - w_2N_2) \quad (2)$$

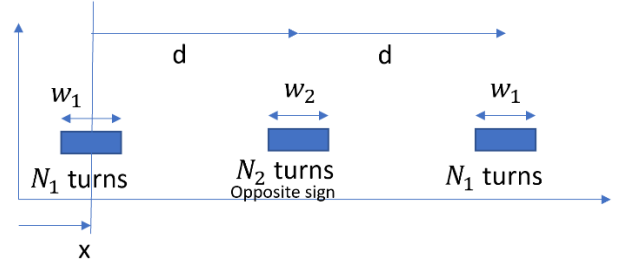

Fig. 4. Effective winding representation for quadrupole bucking determination.

Eqn. (2) indicates that the quadrupole sensitivity ($K_2$) is zero (i.e. the quadrupole is bucked) if

$$2w_1N_1 = w_2N_2 \quad (3)$$

Note that (3) is independent of $x$ as well as $d$. Therefore, for quadrupole bucking, if $w_2 = w_1$, then $N_2 = 2N_1$, as was the case for Fig. 3.

## III. SEXTUPOLE BUCKING PCB CIRCUITS

DQS-bucking can readily be achieved by combining two DQBuck circuits of the previous sections. Besides more straightforward construction, an additional advantage of bucking all harmonics lower than the fundamental is that DQS bucking allows a high-accuracy in-situ calibration of the PCB radial and transverse offset during probe rotation to be performed [4].

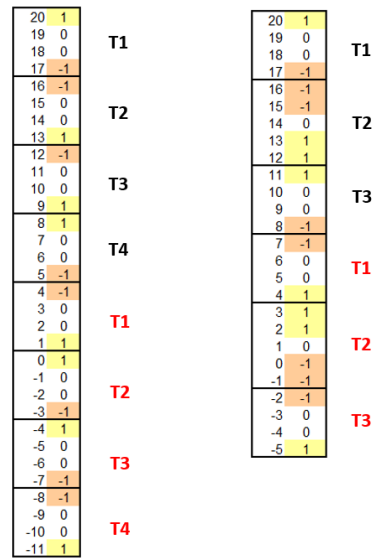

Fig. 5. DQSBuck configurations from combined standard 4-winding DQB (left), and compact DQB (right).

To combine two DQBuck circuits to form a DQSBuck probe, one can simply append another standard DQB circuit of the type from Fig. 1, except with opposite polarity, in a different radial position on the same PCB layer. Two of the 'compact' type DQBuck circuits (Fig. 3) can also simply be connected this way, and achieve an even higher sensitivity for given radial space. Sample layouts for these two combinations are shown in Fig. 5. A third way involves 'interleaving' two compact DQBuck circuits.

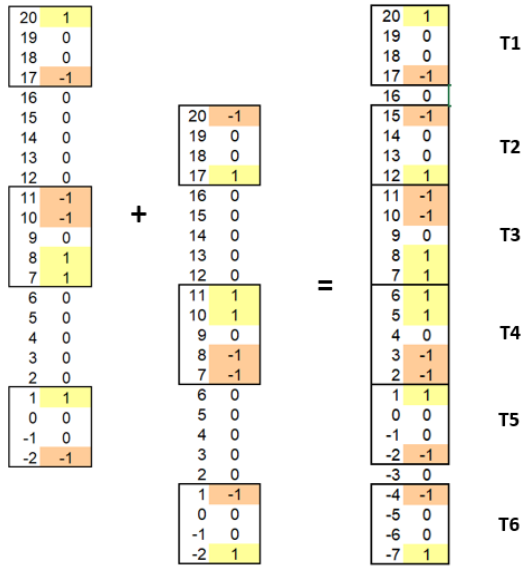

**Fig. 6.** Combination of two 'compact' DQBuck probes to form a DQSBuck probe.

Fig. 6 shows how this combination of compact DQB circuits with opposite polarity are merged to form an interleaved DQSBuck circuit. Note that the individual DQBuck circuits are those of Fig. 3 but with their '$d$' dimension (from Fig. 4) stretched to make space for the combination.

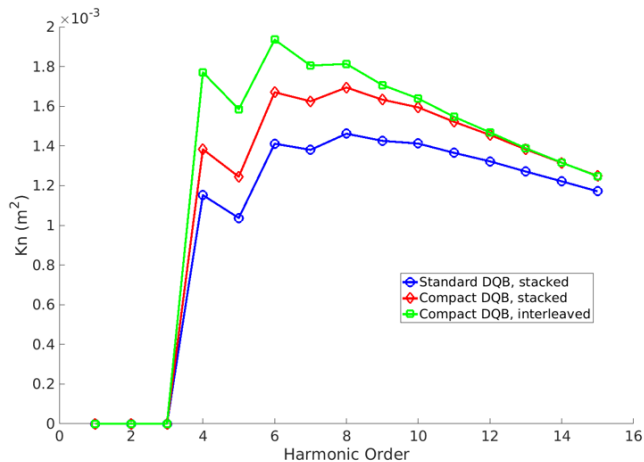

**Fig. 7.** Comparison of DQSB Sensitivity in various bucking configurations.

A comparison of sensitivities between these three types of DQS buck configurations is shown in Fig. 7. As an example, all three configurations are using 78% of the probe diameter for traces, which tends to increase the n=4 harmonic sensitivity at the expense of the n=5. For highest orders, the three configurations tend to converge because the sensitivity is primarily driven by the outermost wire, which has the same radius for all three. It should be noted that though the difference in sensitivity is sizable, gains may fall short of what could be added with additional layers. Of course, the trade-offs might be cost, higher resistance, or for very small probes, not being able to add thickness without actually reducing sensitivity, since thick PCB coils effectively become more 'tangential'. Generally, the pros/cons of increasing layers should be carefully examined for any given PCB design.

An example of an interleaved DQSBuck configuration is the 22.7 mm diameter probe with 14 layer PCB fabricated for the APSU [5] and shown in horizontal layout in Fig. 8.

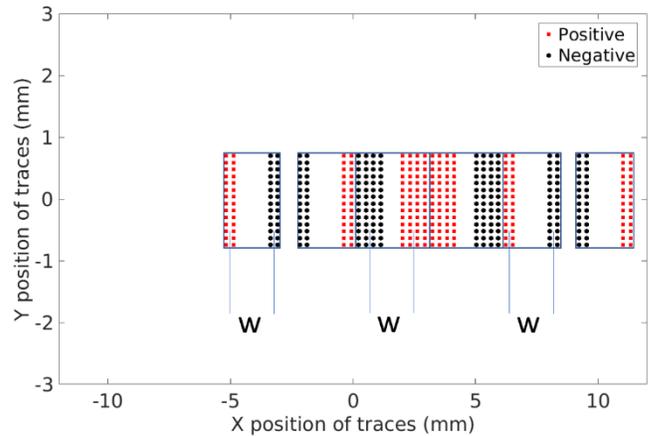

**Fig. 8.** Cross-section of interleaved DQSBuck probe for APSU with uniform centroid widths in central and flanking sections.

Since Eqn. (3) gives a general condition for bucking, we can also try to play off the widths of the DQBuck circuits in Fig. 3 to achieve the bucking. This was done for a 12 mm diameter probe made for NSLSII upgrade measurements [6] having an interleaved 16-layer DQSBuck PCB. This had 4 turns in the central windings (with centroid width 1.35mm) and 3 in the flanking ones, which had centroid widths of 0.9mm (Fig. 9).

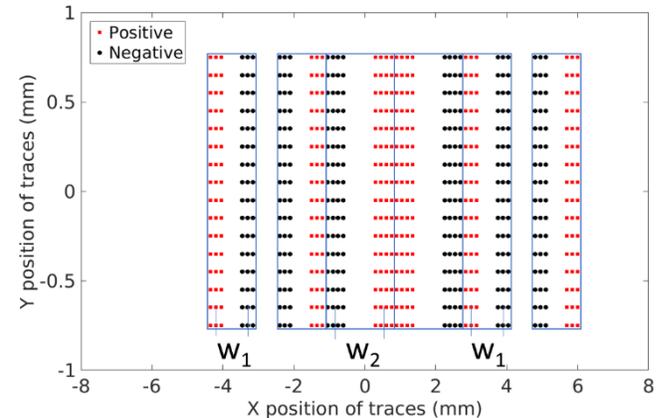

**Fig. 9.** Interleaved DQSBuck PCB cross-section with different centroid widths ($w_1$, $w_2$) in central and flanking sections.

The bucking condition of (3) was thus satisfied since

$$2 * 0.9mm * 3 = 1.35mm * 4$$

Note that for both these DQSBuck probes, the windings occupy about 70% of the diameter of the probe, which is typical for optimal sensitivities (the rotation center is X=0, Y=0 in the plots).

## IV. DQS BUCKING TEST

The 12 mm - diameter probe fabricated for the NSLSIIu measurements (cross-section Fig. 9) is shown in Figure 10. The carbon-fiber reinforced construction should be noted – probe stiffness is crucial for enabling the probe data to be analyzed ideally. However, torsional and transverse vibrations of the whole probe structure are effectively mitigated by the bucking.

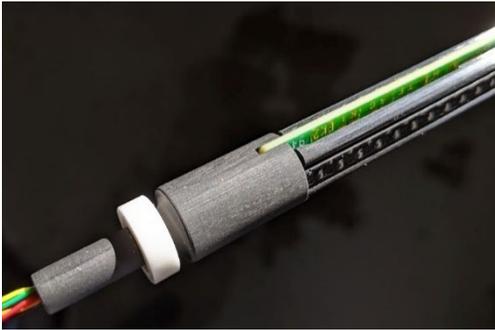

**Fig. 10.** DQSBuck probe with 12mm diameter, 16-layer PCB

A station was built to test the probe, with a small BNL quadrupole magnet that could slide along the probe's 270 mm active length and check probe uniformity (Fig. 11).

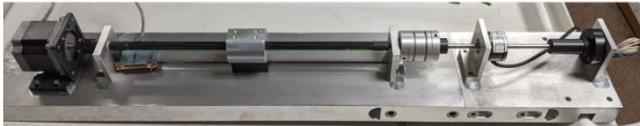

**Fig. 11.** Test fixture for DQSBuck probe with small translatable quadrupole.

The probe radial and transverse offsets are shown in Figures 12 and 13, calculated according to [4]. The uniformity of the probe in both offsets is on the order of 10 microns. The bucking ratio for DQS is shown in Fig. 14, and is above 100 in all positions and rotation speeds. Since the magnet was a quadrupole, with only about 60 units of sextupole, the bucking in a sextupole magnet is likely to be even better, and here just limited because the signal sizes are so small.

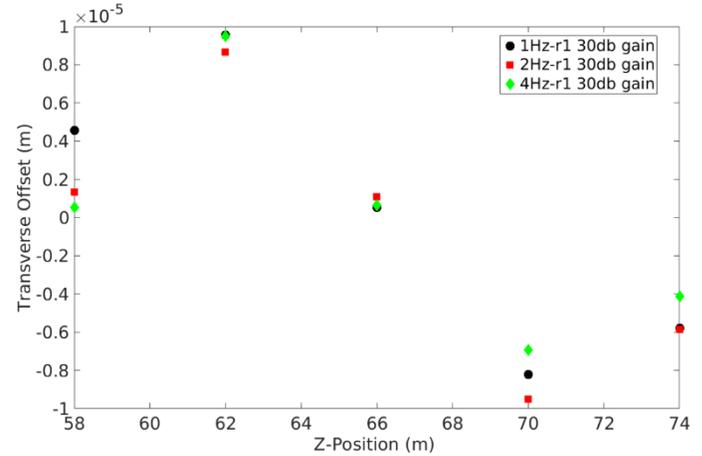

**Fig. 13.** Transverse PCB offset along length of probe.

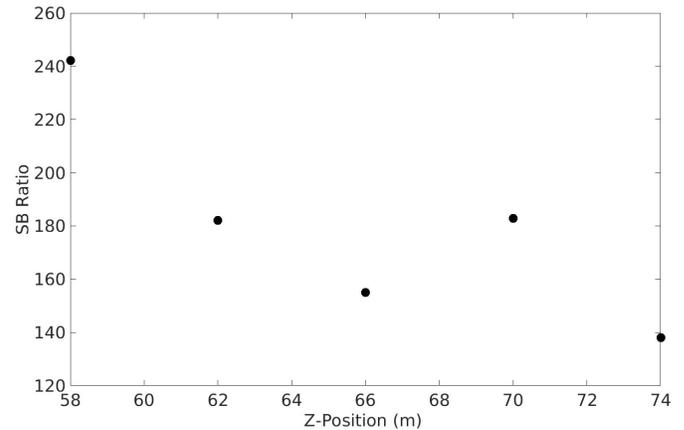

**Fig. 14.** Sextupole bucking ratio along length of DQSBuck probe.

## V. CONCLUSION

The process by which PCBs with simultaneous radial bucking of dipole, quadrupole and sextupole fields can be designed has been described. Good sensitivity and bucking can be obtained, and has been demonstrated in actual probes. A DQSBuck probe composed of interleaved compact DQBuck circuits seems to provide the best sensitivity for available radial space.

## ACKNOWLEDGMENT

The author gratefully acknowledges the significant contributions of D. Assell and N. Moibenko in this work.

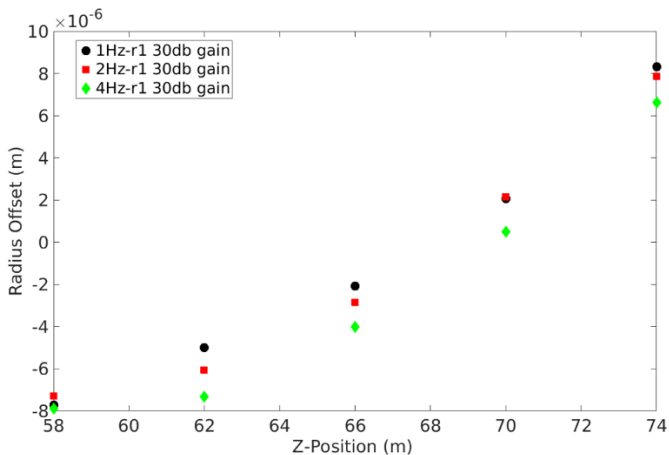

**Fig. 12.** Radial PCB offset along length of DQSBuck probe.